\documentclass{article}

\usepackage{bm}
\usepackage{graphicx}
\usepackage{verbatim}
\usepackage{amsmath}

\usepackage{listings}
\newcommand{\be}{\begin{equation}}
\newcommand{\ee}{\end{equation}}
\newcommand{\ba}{\begin{eqnarray}}
\newcommand{\ea}{\end{eqnarray}}

\newcommand{\la}{\left\langle}
\newcommand{\ra}{\right\rangle}

\newcommand{\pds}{\ensuremath{P_{\delta S}}}
\newcommand{\pdt}{\ensuremath{P_{\delta T}}}

\newcommand{\ds}{{\ensuremath{\delta S}}}
\newcommand{\dt}{{\ensuremath{\delta T}}}
\newcommand{\fs}{\ensuremath{f_S}}
\newcommand{\ft}{\ensuremath{f_T}}
\newcommand{\gs}{\ensuremath{g_S}}
\newcommand{\gt}{\ensuremath{g_T}}

\newcommand{\tf}{\ensuremath{{t_\mathrm{f}}}}

\newcommand{\labeq}[1]{\label{eq:#1}}
\newcommand{\refeq}[1]{(\ref{eq:#1})}

\newcommand{\labsec}[1]{\label{sec:#1}}
\newcommand{\refsec}[1]{(\ref{sec:#1})}

\newcommand{\labfig}[1]{\label{fig:#1}}
\newcommand{\reffig}[1]{(\ref{fig:#1})}

\newcommand{\NOCODE}{}

\title{Path Integral Approach to Uncertainties in SIR-type Systems}
\author{Steven Gratton\thanks{email: stg20@cam.ac.uk}\\ Kavli Institute for Cosmology, University of Cambridge, UK}
\date{June 2, 2020}

\begin{document}
\maketitle

\begin{abstract}
In this paper I show how path integral techniques can be used to put measures on histories 
in ``susceptible-infectious-recovered" (SIR)-type systems.  The standard SIR solution emerges as the 
classical saddle point of the action describing the measure.  One can then expand perturbatively around the
background solution, and this paper goes on to work out the covariance of fluctuations around the 
background solution. Using a Green's function type 
approach, one simply needs to solve additional
ordinary differential equations; an explicit matrix inversion is not required.  The computed
covariance matrix should be useful in the construction of fast likelihoods for fitting the parameters of 
SIR-type models to data.  A comparison of the predictions of the approach to an ensemble of simulations 
is presented.  
\end{abstract}

\section{Introduction}
\labsec{intro}

In understanding the spread of infectious diseases throughout a population, broadly speaking one can either
take a continuous approach and solve deterministic differential equations or else take a stochastic simulation-based 
approach and computes many Monte Carlo realizations of the epidemic.   A disadvantage of the first option is that it gives no indication
of the properties of the uncertainties one should expect to see about the solution, while a disadvantage of the second
option is that it requires the potentially expensive numerical computation of many realizations in order to fully understand 
the behaviour.  In this note I present a scheme that walks a middle ground between these two options, a deterministic procedure that allows 
for the efficient computation of the properties of uncertainties around the mean history.

This paper will treat the simple ``susceptible-infectious-recovered" (SIR) model \cite{sir1927} but the technique is applicable
to more complicated models.  The basic idea is to start by formulating a suitable discrete-time stochastic 
generalization of the SIR model, allowing for appropriate fluctuations in the numbers being infected and recovering. 
One then writes down an approximation to how unlikely some putative ``history" would be in the stochastic framework.
Taking the continuum limit in time, one then obtains a ``measure" of the likelihood of that history occurring if the 
model is true.  This measure is written as the exponential of minus the ``action'' of the history (following physics naming
conventions in mechanics and quantum theory).  

Once one has an action, one has a probability distribution on entire histories of the evolution of the system under consideration
and so has access to all possible information about the statistics of such histories.  This is an ``imaginary time'' version of Feynman's  
path integral for quantum systems \cite{Feynman:1965:QMP}.  As in mechanics, the saddle point of the action determines the most likely history to occur.  Then,
as in quantum mechanics, by considering deviations away from the saddle point solution, one can work out the statistical properties of
the deviations of the variables such as their covariances at both equal and unequal times.  

Actually, it turns out that one need not explicitly evaluate the path integral over the fluctuations in order to obtain the unequal time
correlators (or UETCs) in the case where the fluctuations are small; rather one can obtain them from the form of the expansion of the
action to second order about the saddle point history.  This effectively gives one an inverse covariance matrix, whose inverse can be
obtained without a matrix inverse operation via a multi-variable version of the ``Green's function'' technique.   It is shown that this
only requires solutions of the classical Hamiltonian equations for the fluctuation variables. 

So, simply by solving ordinary differential equations, one can deterministically obtain both the typical behaviour of disease spread in 
SIR-type models and the properties of the uncertainties  around that spread.

Analogous path integral techniques have been found useful by the author and colleagues in understanding the properties 
of quantum fluctuations in a variety of situations in models of the early Universe; see e.g. \cite{Gratton:1999ya,Gratton:2000fj,Gratton:2001gw,
Gratton:2010zv}.  The work of \cite{2020arXiv200509625A} also discusses fluctuations around SIR-type models, from an alternative viewpoint, and
provides references to the literature on infectious disease modelling and inference.   

Section \refsec{model} describes the model situation, and Sec.\ \refsec{acthist} constructs the action weighting each history in the path integral approach.  Sections \refsec{sadfluc} and \refsec{covmat} then study saddle points and fluctuations around them, showing how to compute covariance matrices for quantities of interest.  
Section \refsec{numres} presents a comparison of these results with numerical simulations, and a discussion and conclusions are given in Sec.\ \refsec{discon}.  
\ifdefined\NOCODE
\else
An appendix presents a python implementation of the method.
\fi

\section{Model Construction}
\labsec{model}

Consider a population of total number $N$, with $S(t)$ susceptible people and $I(t)$ infectious people at time $t$.  Let us imagine that every day an infectious
person meets with $d$ other people, and infects each of them with probability $\mu$ if they have not already been infected.  Overnight each infectious
person (not including those just infected that day) recovers with a probability $\gamma$.

So from day to day we would expect
\ba
\la \Delta S \ra&=&-\mu d \, I \frac{S}{N}, \labeq{ds}\\
\la \Delta I \ra&=&\mu d \, I \frac{S}{N} - \gamma I. \labeq{di}
\ea
The $S/N$ factor in the first equation accounts for the fact that only susceptible people can be infected (assuming one cannot catch the disease
twice for now) and assumes that a typical infectious person is surrounded by typical non-infectious people.

What fluctuations around these means might we expect?  In terms of infections, we have a total of $d \, I S$ infection events a day, which we may take to
be independent.  So we expect a binomial distribution of successes, with the number of trials being $d \, I $ and the probability of success per trial being $  
\mu S / N$.  This indeed has mean $ \mu d \, I S /N$, and has variance $ \mu d \, I S /N (1-\mu)$.   

The number of recoveries might also be binomially distributed, and we shall consider this case here.  There will be $I$ trials per day with a probability of success per trial being $\gamma$, giving mean $\gamma I$ and variance $\gamma I (1-\gamma)$. 

To match convention we can denote $\mu d$ by $\beta$.  In the context of COVID-19, if people are typically infectious for about five days say, then 
$\gamma$ will be around $1/5$.  For the disease to be able to spread initially (when $S \approx N$), from Eq.\ \refeq{di} we see need $\beta - \gamma>0$, or the \emph{reproduction number} $R_0 \equiv \beta / \gamma$ to be greater than unity.  COVID-19 seems to typically spread with $R\sim 2.5$, so $\beta \sim 0.5$.  If $d$ is of order ten say, then $\mu$ must be around $0.05$.  Hence we are close to being in the Poisson limit for the number of daily infections, with mean and
variance then both $\beta I S/N$.  For the number of recoveries, it is not so clear that we are in the Poisson limit but for simplicity we shall assume this for now.  If things are not changing too quickly per day, we can treat a few, $\Delta t$, days together at a time.  We are then just effectively multiplying the number of potential events by $\Delta t$.  Keeping $\Delta t$s around will also presently help us in taking the continuum limit.

\section{Histories, Actions and Path Integrals}
\labsec{acthist}

From Eqs.\ \refeq{ds} and \refeq{di} and our discussion above, we see that the behaviour of $S$ and $I$ are coupled.  To disentangle this, we introduce a new
variable $T\equiv S+I$.  Now, consider a possible ``history" of the disease spread, corresponding to values for $S$ and $T$ for all times $t$ (in days).
Then, over a few $\Delta t$ days, focussing for the moment on $S$, the difference between the change in $S$  in this history and the expected change $\Delta t \cdot \la \Delta S(t) \ra$, coming from Eq.\ \refeq{ds}, must be put down to 
fluctuations.  Approximating the Poisson by a gaussian with the corresponding mean $-\beta I (S/N) \Delta t$ (negative since an infection reduces $S$) and variance $\beta I (S/N) \Delta t$, the probability density of such a fluctuation is then given by
\be
\frac{1}{\sqrt{2 \pi \beta I (S/N) \Delta t}}
\exp \left(-\frac{1}{2}\frac{ (S(t+\Delta t) - S(t)+\beta I (S/N) \Delta t)^2}{\beta I (S/N) \Delta t}\right).
\ee 
Note how the $\Delta T$ factors in the exponent can be rearranged to write the exponent as
\be
-\frac{1}{2}\frac{\left(\frac{S(t+\Delta t) - S(t)}{\Delta t}+\beta I (S/N)\right)^2}{\beta I (S/N) } \Delta t.
\ee
Multiplying the probability distributions from the groups of time steps making up the entire history, we obtain an exponent
\be
-\frac{1}{2}\sum \frac{\left(\frac{S(t+\Delta t) - S(t)}{\Delta t}+\beta I (S/N)\right)^2}{\beta I (S/N) } \Delta t.
\ee
Now we may take the continuum limit $\Delta t\rightarrow0$ and thus obtain a probability density or ``measure" $e^{-S_S}$  with ``action" $S_S$ given by
\be
S_S=\int_0^{t_\mathrm{f}} dt \frac{1}{2} \frac{(\dot{S}+f)^2}{f},
\ee 
where we have defined $f \equiv \beta I (S/N)$ for notational simplicity (and we have neglected any history dependence of the prefactor), assumed initial and final times of $0$ and $t_\mathrm{f}$ respectively and denoted time derivatives by an overdot.  Following a similar
argument through for $T$ we obtain an action for $T$ of 
\be
 S_T=\int_0^{t_\mathrm{f}}dt \frac{1}{2} \frac{(\dot{T}+g)^2}{g}. 
 \ee
with $g\equiv \gamma (T-S)$ (since $I=T-S$).  

Hence our full measure on a history is given by an action
\be
S_\mathrm{E}=\frac{1}{2} \int_0^{t_\mathrm{f}} dt \, \left( \frac{(\dot{S}+f)^2}{f}+\frac{(\dot{T}+g)^2}{g} \right). \labeq{se}
\ee
(Here we have followed a physics convention and used the subscript $_\mathrm{E}$  to denote a ``Euclidean", as opposed to ``Lorentzian", action.)  Had 
we not introduced $T$ we could have introduced a 2-dimensional correlated gaussian for changes in $S$ and $I$ at each time step and ended up with an equivalent result. 

With a measure, we can obtain the expectation value of some quantity $\cdots$ with:
\be
\langle \cdots \rangle = \frac{\int Dq \cdots e^{-S_\text{E}}}{\int Dq\, e^{-S_\text{E}}}. \labeq{av}
\ee
Here $Dq$ denotes the functional path integral over the variables collectively written together as $q$. One considers integrating over all paths subject to appropriate boundary
conditions.  If one knew the system starts in some state for example, then one would constrain all histories to start in that manner.  One can also impose final state
conditions if one is interested in knowing what might have happened in between two known states, or leave the final conditions free.  Else evaluating the path integral as a function of final state (or equivalently choosing $\cdots = \delta(q(\tf)-q_\text{f})$) will give the probability for the final state $q_\text{f}$.

Some common choices for $\cdots$ are one of the variables at some time $t_1$, e.g.\ $S(t_1)$ or say the squares of a variable at equal times (e.g.\ $S(t_1)^2$).  From these the variance of $S$ at $t_1$ can be derived.  More generally we can
work out the covariance between two variables at different times, $\langle \langle q_i (t_1) q_j (t_2) \rangle \rangle$, also known as an unequal time correlator (UETC).

\section{Saddle Points and Fluctuations}
\labsec{sadfluc}

Equation \refeq{av} can in principle be used to compute all quantities of interest within a model.  However the functional integral cannot usually be evaluated.  Instead
a saddle point approximation can be performed.  Here one finds the history that minimizes the action, and considers histories around this one.

The Euler-Lagrange equations for varying $S_\text{E}$ with respect to $S$ and $T$ yield
\ba
\frac{d}{dt}\left(\frac{\dot{S}+f}{f}\right) &=& \frac{(\dot{S}+f)f_S}{f}-\frac{(\dot{S}+f)^2 f_S}{2 f^2}+\frac{(\dot{T}+g)g_S}{g}-\frac{(\dot{T}+g)^2 g_S}{2 g^2}, \labeq{els}\\
\frac{d}{dt}\left(\frac{\dot{T}+g}{g}\right) &=& \frac{(\dot{S}+f)f_S}{f}-\frac{(\dot{S}+f)^2 f_T}{2 f^2}+\frac{(\dot{T}+g)g_T}{g}-\frac{(\dot{T}+g)^2 g_T}{2 g^2}. \labeq{elt}
\ea
Here subscripts $_S$ and $_T$ denote partial derivatives with respect to $S$ and $T$.   We see that if $\dot{S}+f=0$ and $\dot{T}+g=0$ holds at some time, then it must actually hold at all times.  But $\dot{S}+f=0$ and $\dot{T}+g=0$ are just the standard SIR equations, so we have seen how their solutions can emerge as saddle points of the action in the path integral approach.

We now go on to argue that the SIR-type solutions are actually the most likely saddle point if the imposed boundary conditions allow.  As discussed above, the probability for obtaining the final state $q_\text{f}$ given some initial state $q_\text{i}$ is given by evaluating the path integral over all paths that interpolate between $q_\text{i}$ and $q_\text{f}$.  In the saddle point approximation, to leading order this is just $e^{-S_\mathrm{E}(q_\text{f},q_\text{i})}$, where $S_\mathrm{E}(q_\text{f},q_\text{i})$ is the action calculated for the interpolating path.  The interpolating path's initial velocities will be chosen to make the $q(t_\text{f})$ under evolution of Eqs.\ \refeq{els} and \refeq{elt} indeed be the desired $q_\text{f}$.   
By inspection of Eq.\ \refeq{se}, we see that the history that has the $S(t_\text{f})$ and $T(t_\text{f})$ given by integrating $\dot{S}+f=0$ and $\dot{T}+g=0$ from the initial $S(0)$ and $T(0)$ must have an action of $0$.  Then, as the action is positive semi-definite, all other paths, which have different field vales at the final time, must have higher action and so be less likely.

Having found the most likely path or background solution for $S(t)$ and $T(t)$, we now consider fluctuations around it, expanding $S_\text{E}$ to second order in $\delta S(t)$ and $\delta T(t)$.  The form of the action and the 
most likely path having  $\dot{S}+f=0$ and $\dot{T}+g=0$ make this particularly straightforward, yielding
\be
S_2=\frac{1}{2} \int_0^{t_\mathrm{f}} dt  \frac{(\dot{\ds}+\fs \ds +\ft \dt)^2}{f}+\frac{(\dot{\dt}+\gs \ds +\gt \dt)^2}{g}. \labeq{s2}
\ee
 
\section{Covariance Matrix Construction}
\labsec{covmat}

Evaluating $e^{-S_2}$, with $S_2$ from Eq.\ \refeq{s2}, for any given $\delta S(t)$ and $\delta T(t)$ will tell us how much more unlikely such a history is than the standard SIR solution.  Being quadratic in the fluctuations,  $S_2$ can productively thought of as minus the exponent of a multi-dimensional gaussian probability distribution in the limit as the number of dimensions tends to infinity as the duration of the time step tends to zero.  In fact, let us pass back to discrete time for clarity in the following.  (Note now though $\Delta t$ is no longer being thought of as a multiple number of days, just some non-zero fraction of a day.)  In matrix notation, $-2 S_2$ corresponds to
\be
\Delta t
\begin{pmatrix}
\ds\\
\dt
\end{pmatrix}^{\text{T}}
C^{-1}
\begin{pmatrix}
\ds\\
\dt
\end{pmatrix} \labeq{s2mat}
\ee
where \ds\ and \dt\ are now viewed as column vectors with each row corresponding to a timestep, $f$, $g$ and their partial derivatives are now viewed as appropriate (diagonal) matrices, superscript ${}^\text{T}$ denotes matrix transpose, $C^{-1}$ (so-named by analogy to the inverse covariance matrix for a gaussian distribution) is the matrix
\be
\begin{pmatrix}
(D+\fs)^\text{T} f^{-1} (D+\fs)+\gt^\text{T} g^{-1} \gt&(D+\fs)^\text{T} f^{-1}\ft + \gs^\text{T} g^{-1} (D+\gt) \\
\ft^\text{T} f^{-1} (D+\fs)+(D+\gt)^\text{T} g^{-1}\gs  & (D+\gt)^\text{T} g^{-1} (D+\gt)+\ft^\text{T} f^{-1} \ft
\end{pmatrix} \labeq{cinv}
\ee
and $D$ is the appropriate discrete-time time-derivative operator:
\be
D\equiv \frac{1}{\Delta t}
\begin{pmatrix}
1 & 0 & 0 &\cdots&0\\
-1 & 1 & 0 &\cdots&0\\
\vdots&&&&\\
 \cdots &0&-1&1&0\\
\cdots& 0 & 0 & -1 &1\\
\end{pmatrix}. \labeq{ddef}
\ee
It is important at this stage to consider the sorts of boundary conditions we shall wish to impose.  For the problem as formulated here, we suppose were are given
the initial values for $S$ and $T$.  So $\ds(0)$ and $\dt(0)$ should initially be zero and any variation of them should not be considered.  Hence our vectors for \ds\ and \dt\ start from the first timestep with $\ds(\Delta t)$ and $\dt(\Delta t)$.  By the construction of the path integral, we see it involves ``forward" differences and so we have correspondingly constructed our $D$ matrix not to include anything from a step ``beyond" the end of \tf.  We need not impose any conditions on the \ds\ and \dt\ at the final timestep; indeed we often want to determine what these should be at the end of the evolution.  We shall see presently the role of the potentially-dangerous unbalanced $1/\Delta t$ in the first row of $D$ (and of the unbalanced  $1/\Delta t$ in the last row of $D^\text{T}$) in ensuring we find the correct solution.  (Unlike for $D$, we can allow $f$, $g$ and their partial derivatives to ``overrun'' as required to give invertible operators with vanishing error in the continuum limit.)

If the number of timesteps is not too large, one could feasibly directly calculate the inverse of $C^{-1}$, which, when
divided by $\Delta t$, would directly give the covariance matrix for the elements of \ds\ and \dt.   In any case, we can proceed without having to explicitly perform the 
matrix inverse as follows.  We want to find $G$, the matrix that satisfies:
\be
C^{-1}
\cdot G
=
\frac{1}{\Delta t} I
\ee
where $I$ is the appropriately-sized identity matrix.  Now, consider a matrix with twice as many rows and columns as $C^{-1}$, written in block form as:
\be
\begin{pmatrix}
-a & b \\
b^\text{T} &c 
\end{pmatrix}. \labeq{matex}
\ee
The lower-right component of its inverse is just $ (c+b^\text{T} a^{-1} b)^{-1}$ (see e.g.\ \cite{numrec}).  Reversing the argument
we see that the inverse of a matrix that can be written in the form $c+b^\text{T} a^{-1} b$ is just the lower-right block of the inverse of the larger matrix
 \refeq{matex}.   Now, \refeq{cinv} can be written as:
 \be
 C^{-1} =
\begin{pmatrix}
D+\fs & \ft \\
\gs & D+\gt
\end{pmatrix}^\text{T}
\begin{pmatrix}
f^{-1}&0 \\
0 & g^{-1}
\end{pmatrix}
\begin{pmatrix}
D+\fs & \ft \\
\gs & D+\gt
\end{pmatrix}
\ee
and so $G$ must just be the lower-right block of the matrix $K$, where 
\be
B K = \frac{1}{\Delta t} I, \labeq{bkeq}
\ee
and $B$ is the matrix
\be
B \equiv
\begin{pmatrix}
-f&0&(D+f_S)&f_T \\
0&-g&g_S&(D+g_T) \\
(D+f_S)^\text{T}&g_S^\text{T}&0&0 \\
f_T^\text{T}&(D+g_T)^\text{T}&0&0
\end{pmatrix}.
\ee
By construction, $D$ is a discrete-time time-derivative operator acting on column vectors to its right.  So $D^\text{T}$ must act as
the time-derivative operator on row vectors to its left.  But, away from the ends, by inspection of \refeq{ddef},  $D^\text{T}$ also acts as \emph{minus} a time-derivative
operator on column vectors to its right, which we can emphasize by defining $D'\equiv -D^\text{T}$.  
$B$ is very sparse, only mixing variables amongst themselves at equal or neighbouring times.   
Consider for a moment a column on the right hand side of $K$. Writing it as a stack of vectors $P_\ds$, $P_\dt$, $\ds$ and $\dt$, from Eq.\ \refeq{bkeq} it must satisfy:
\ba
-f \pds+(D+f_S) \ds +f_T \dt&=&0, \labeq{finiteds}\\
-g \pdt +g_S \ds +(D+g_T) \dt &=& 0, \labeq{finitedt}\\
(-D'+f_S) \pds + g_S \pdt &=& \delta_{t,t_2} /\Delta t \text{ or } 0, \labeq{finitepds} \\
f_T \pds +(-D'+g_T) \pdt &=& 0 \text{ or } \delta_{t,t_2}/\Delta t, \labeq{finitepdt}
\ea
where the delta function appears in one equation or the other and at the particular timestep $t_2$ depending on the precise column considered.    Apart from the few rows with ``unbalanced'' $1/\Delta t$'s in them from the derivative operator, we are now in 
a position to take the continuum limit $\Delta t \rightarrow 0$ again to find that the column must satisfy:
\ba
-f \pds+\dot{\ds}+f_S \ds +f_T \dt &=&0, \nonumber \\
-g \pdt +g_S \ds +\dot{\dt} + g_T \dt &=& 0, \nonumber\\
-\dot{\pds}+f_S \pds + g_S \pdt &=& \delta(t-t_2) \text{ or } 0, \nonumber \\
f_T \pds -\dot{\pdt}+g_T \pdt &=& 0 \text{ or } \delta(t-t_2). \labeq{hamgf}
\ea
Neglecting the delta functions on the right hand side, Hamilton's equations for the action $S_2$ (Eq.\ \refeq{s2}) have emerged, the momenta serving as auxiliary variables enabling us to straightforwardly solve for the correlation functions for the degrees of freedom \ds\ and \dt.  Explicitly, varying $S_2$ with respect to 
$\dot{\ds}$ and $\dot{\dt}$ and introducing the Hamiltonian $H$ as usual in classical mechanics, one finds $\pds=(\dot{\ds}+\fs \ds +\ft \dt)/f$, $\pdt=(\dot{\dt}+\gs \ds +\gt \dt)/g$, with
\be
H=\frac{f}{2} \pds^2 + \frac{g}{2} \pdt^2 - \pds (\fs \ds + \ft \dt) - \pdt ( \gs \ds + \gt \dt).
\ee
Hamilton's equations then just read:
\ba
-f \pds+\dot{\ds}+f_S \ds +f_T \dt &=&0, \labeq{hamds} \\
-g \pdt +g_S \ds +\dot{\dt} + g_T \dt &=& 0, \labeq{hamdt}\\
-\dot{\pds}+f_S \pds + g_S \pdt &=& 0, \labeq{hampds}\\
f_T \pds -\dot{\pdt}+g_T \pdt &=& 0, \labeq{hampdt}
\ea
to compare to \refeq{hamgf}.\footnote{The appearance of momenta and Hamilton's equations in this manner is actually quite subtle because $B$ unlike $C^{-1}$ is not positive-definite, corresponding to $H$ not being bounded below.  Thus, unlike in quantum mechanics, we cannot introduce simply introduce momenta at an early stage in the path integral, at least if we wish to keep the momenta real and not let them go into the complex plane.  If we did introduce real momenta, we would have to think about finding the extremum of the extended system rather than integrating over all configurations.}

Equations \refeq{hamgf} constitute a two-dimensional generalization of a Green's function.   For the elements $\langle \langle \ds(t_1) \ds(t_2) \rangle \rangle$
and $\langle \langle \dt(t_1) \ds(t_2) \rangle \rangle$, considered as a function of $t_1$, we need to patch together a solution to Hamilton's equations for $t_1<t_2$ and for $t_1>t_2$ such that there is a jump of $-1$ in \pds\ going from $t_1=t_2-\epsilon$ to $t_1=t_2+\epsilon$ (as seen by integrating the third equation of \refeq{hamgf} over $t_1$ from $t_2-\epsilon$ to $t_2+\epsilon$).  Similarly,  the elements $\langle \langle \ds(t_1) \dt(t_2) \rangle \rangle$
and $\langle \langle \dt(t_1) \dt(t_2) \rangle \rangle$ need solutions with a jump of $-1$ in \pdt\ at $t_1=t_2$. 

Let us now think about the ``unbalanced'' rows; we shall see that they naturally provide the appropriate boundary conditions we need.  The first rows of Eqs.\ \refeq{finiteds} and \refeq{finitedt}, containing the terms $\ds(1)/\Delta t$ and $\dt(1)/\Delta t$ respectively (``1'' denoting the first element), tell us that the first elements of \ds\ and \dt\ must tend to zero as $\Delta t \rightarrow 0$, whereas the last rows of Eqs.\ \refeq{finitepds} and \refeq{finitepdt}, containing the terms  $-\pds(N)/\Delta t$ and $-\pdt(N)/\Delta t$ respectively (``$N$'' here denoting the final element), tell us that the last elements of \pds\ and \pdt\ must tend to zero as $\Delta t \rightarrow 0$.  So, in the continuum limit, we need $\ds=\dt=0$ initially, and $\pds=\pdt=0$ finally. 

As Eqs.\ \refeq{hamds}-\refeq{hampdt} are linear, we need not re-solve Hamilton's equations for each value of $t_2$ that we are interested in knowing correlation
functions for; rather  we may just take different linear combinations of an appropriate set of pre-computed solutions to satisfy either the initial or final boundary conditions.  Here, we need four pre-computed solutions, conveniently chosen to each have a different single canonical variable being initially non-zero.
As it happens, in constructing our correlation functions for this problem we are aided by the special circumstance that Eqs.\ \refeq{hampds} and \refeq{hampdt} preserve $\pds=\pdt=0$, so the solutions satisfying the final boundary conditions have $\pds=\pdt=0$ initially also.  

\section{Comparison with Numerical Results}
\labsec{numres}

\ifdefined\NOCODE
The author has written code that computes the Green's functions for fluctuations around an SIR model.  The code also computes a number of stochastic realizations of the situation as discussed in Sec.\ \refsec{model}, allowing for comparison between the analytic and stochastic handling of fluctuations.  
\else
Code that computes the Green's functions for fluctuations around an SIR model is presented in Appendix \ref{sec:pyapp}.  The code also computes a number of stochastic realizations of the situation as discussed in Sec.\ \refsec{model}, allowing for comparison between the analytic and stochastic handling of fluctuations.  
\fi

Parameters assumed for the model are presented in Table \ref{tab:params}.  $N$ has been deliberately chosen to be quite small in order to yield quite large variations.  Indeed, from the form of the action \refeq{se}, we can, up to the discreteness of the initial conditions (i.e.\ keeping $I(0)$ some finite integer greater than zero independent of how large $N$ is), expect the fractional fluctuations to scale as $1/\sqrt{N}$, and this is verified numerically.  In addition, whilst keeping $\gamma$ at a COVID-19-type level, we have lowered $\beta$ somewhat from the unmitigated COVID-19-type level in the description of Sec.\ \refsec{model} to obtain an $R_0$ of $1.5$, in order to better illustrate subtleties associated with epidemics that fail to take off.  Such a value might also be appropriate for modelling COVID-19 dynamics with some intervention schemes in place.

\begin{table}[t!]
\centering
\begin{tabular}{|c|c|}
\hline
Parameter&Value\\
\hline
$\beta$&0.3 /day\\
$\gamma$&0.2 /day\\
$R_0$&1.5\\
$N$&5000\\
$I(0)$&10\\
$R(0)$&27\\
\hline
\end{tabular}
\caption{Table showing the parameters used in the model used in Sec.\ \refsec{numres}.}
\label{tab:params}
\end{table}

\begin{figure}[b!]
\includegraphics[width=\linewidth]{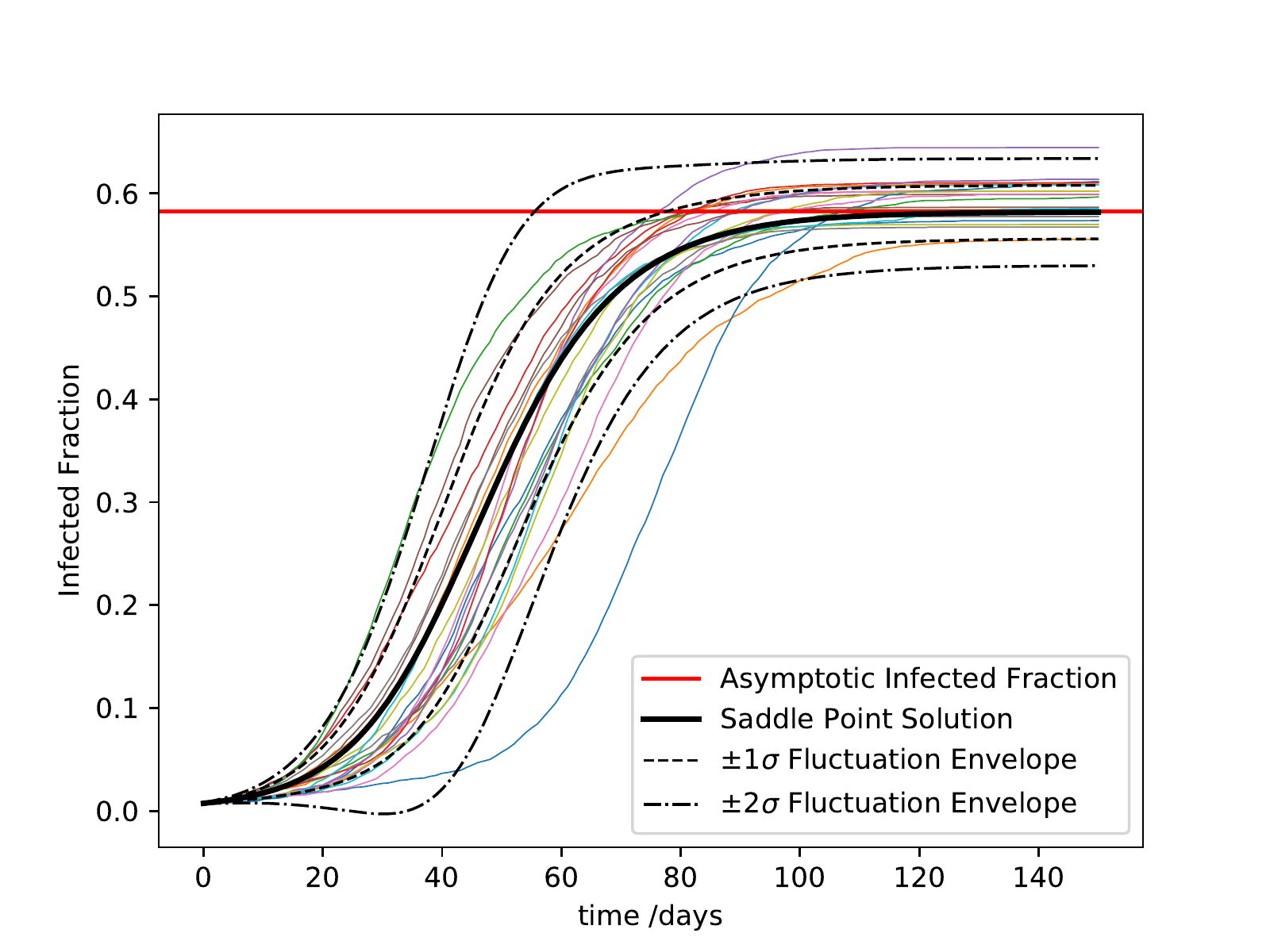}
\caption{Plot illustrating the time evolution of the infected fraction of the population.  The thin coloured lines are twenty stochastic realisations from simulations, while the solid line is the SIR saddle point solution.  The envelopes for fluctuations around the saddle point have been computed using the Green's function technique.}
\label{fig:pathsim}
\end{figure}

To achieve sub-percent level agreement of the simulations with the model it was necessary to allow $S$ and $I$ to evolve about four or more times over the 
course of a single day rather than the once a day described in Sec.\ \refsec{model}.  In particular, the daily updating case saw a final infected fraction that was some
way away from the ``classic" SIR result $r$ coming from solving the transcendental equation $(\gamma/\beta) \log (1-r)+r=0$.
In addition, initial conditions were chosen to roughly agree with the early evolution of an outbreak in an SIR model, in that if there are finite initial numbers of infectious people $I(0)$ at the start, roughly $ R/(R-1)*(I(0)-1)$ people should be taken to have already recovered.

\begin{figure}
\includegraphics[width=\linewidth]{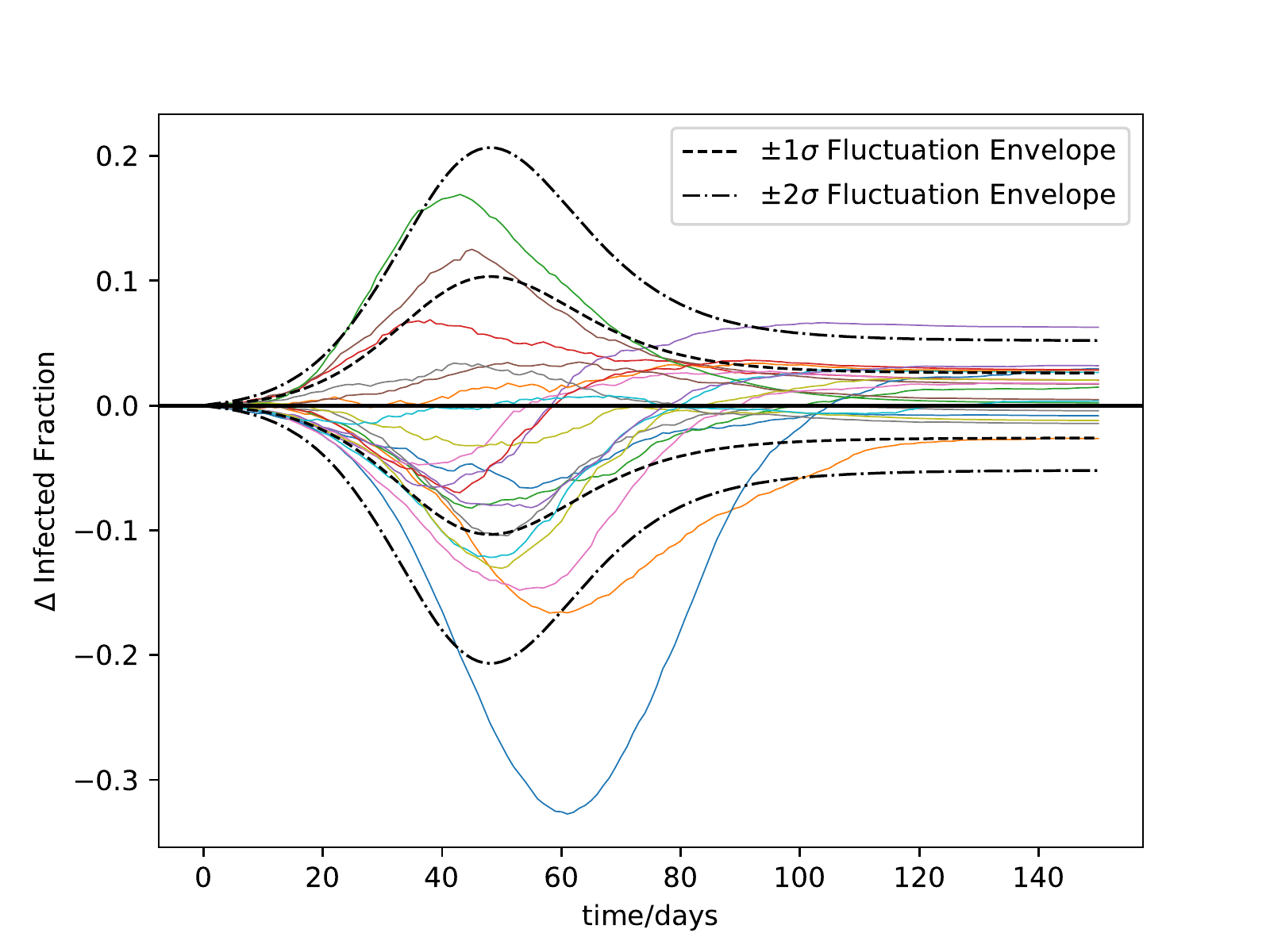}
\caption{As for Fig.\ \reffig{pathsim} but now showing residuals of paths relative to the SIR saddle point.}
\label{fig:pathsimdiff}
\end{figure}

Figures \reffig{pathsim} and \reffig{pathsimdiff} compare twenty realizations of the evolution of the infected fraction from the stochastic simulations to the analytic saddle point solution and $\pm 1\sigma$ and $\pm 2\sigma$ uncertainty envelopes computed using the path integral technique.

Figure \reffig{pathhist} compares the scatter in the final asymptotic infected fraction from 10000 simulation realizations to a gaussian with the computed variance.  The path integral technique evidently performs very well.  However, as foreshadowed by the lower $2\sigma$ contour in Fig.\ \reffig{pathsim} going negative for a time early on, in a minority of cases, roughly 150 in the test shown here, the epidemic never takes off and so there are negative outliers for which the gaussian path integral approximation fails critically for.  The histogram with the outliers removed has standard deviation $0.027$ to be compared with the path integral value of $0.026$ (whereas including the outliers yields a standard deviation of $0.080$). Actually,  Fig.\ \reffig{pathsim} suggests that it might be possible to estimate the fraction of paths that peter out by evaluating the path integral at a relatively early time whilst the gaussian approximation is still reasonable and seeing what fraction of the distribution is below zero. 

\begin{figure}[t!]
\includegraphics[width=\linewidth]{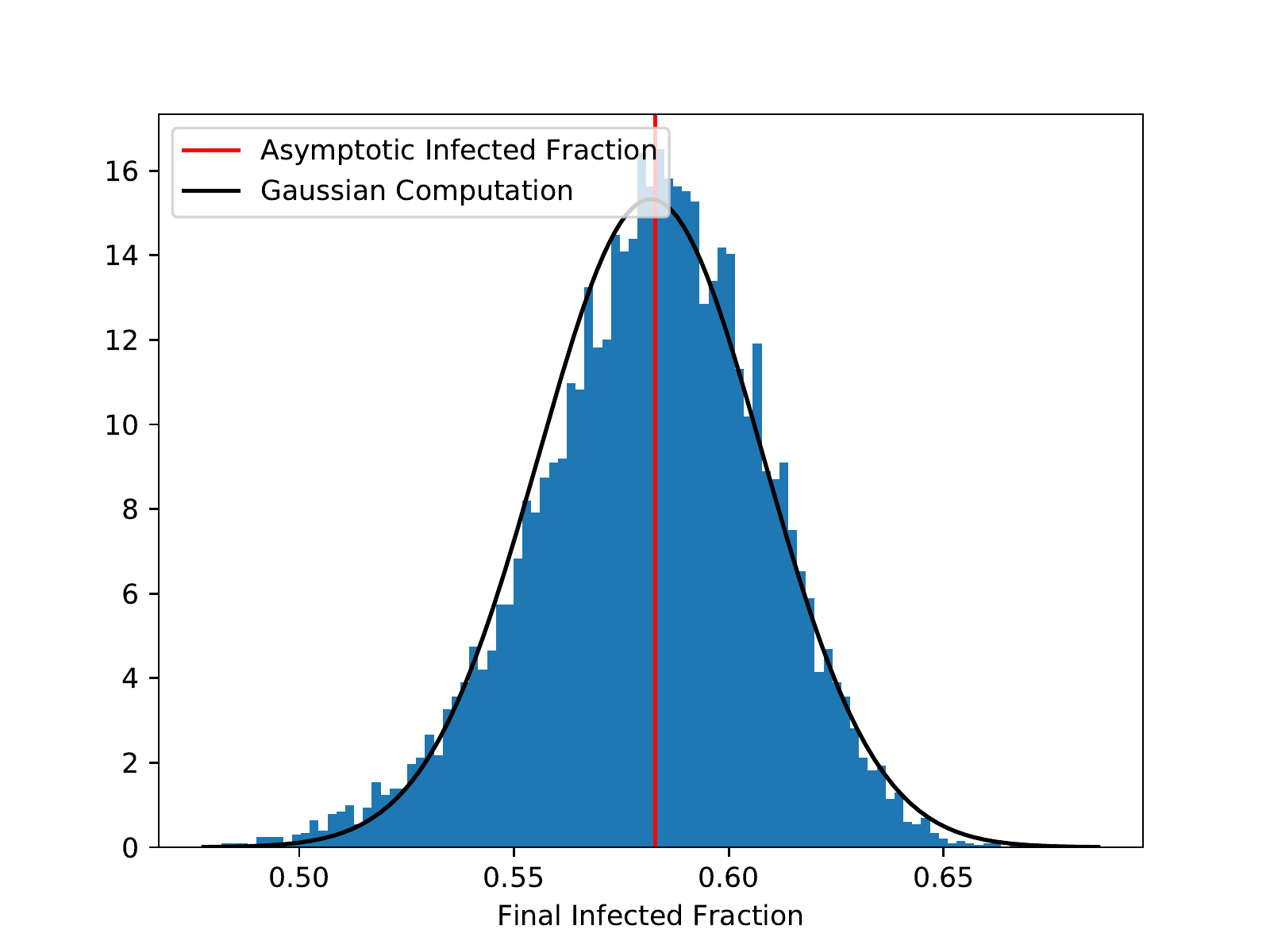}
\caption{Normalized histogram of the final infected fraction from 10000 numerical realizations, compared to a gaussian with mean and variance given by the path integral approach.  The classic SIR asymptotic infected fraction is shown with the red vertical line.  Note that there is a negative tail, not shown, coming from the order 180 simulations for which the epidemic happened to peter out early on.}
\labfig{pathhist}
\end{figure}


Figure \reffig{pathcorr} illustrates the unequal time correlation matrix for $S$, computed by the Green's-function-type procedure discussed above in Sec.\ \refsec{covmat}.
It is apparent that fluctuations away from the saddle point solution are highly correlated in time, as can also be seen from the simulation approach by examining the trajectories in Figs.\ \reffig{pathsim} and \reffig{pathsimdiff}.  This raises the question of whether the trajectories can roughly be matched on to one another by displacing them in time.  The top plot of Fig.\ \reffig{pathdisp} attempts this, and it certainly seems that the evolving portions of all the simulated trajectories are rather similar.
In physics, time derivatives of saddle point solutions often yield near zero-modes of the action (i.e.\ field perturbations that are unsuppressed by the action), due to time translation invariance (broken only by boundary conditions). 
 By inspection of Eqs.\ \refeq{hamds}--\refeq{hampdt}, we see that $\ds=\dot{S}, \dt=\dot{T}$ (with $\pds=\pdt=0$) is indeed a solution of the perturbation equations.   However, as both $\dot{S}$ and $\dot{T}$ are non-zero initially, this is not an actual  zero mode for the case investigated here\footnote{If we were in a situation in which different boundary conditions to $\ds=\dt=0$ were appropriate, corresponding say to describing the middle part of an epidemic without knowing when it started, zero modes might become physically relevant.  Then $C^{-1}$ becomes non-invertible and one would calculate the analogue of a ``projected'' Green's function instead.  This would then describe fluctuations orthogonal to the zero mode, and would be of use in looking at residuals compared to a best-fit background solution.}.  
 Indeed, with the same time shifts as chosen to match $S$, in the bottom plot of Fig.\ \reffig{pathdisp} we see that there is some differentiation in the histories of $I$, which would not be the case were all paths time shifts of one another.

 \begin{figure}[t!]
\includegraphics[width=\linewidth]{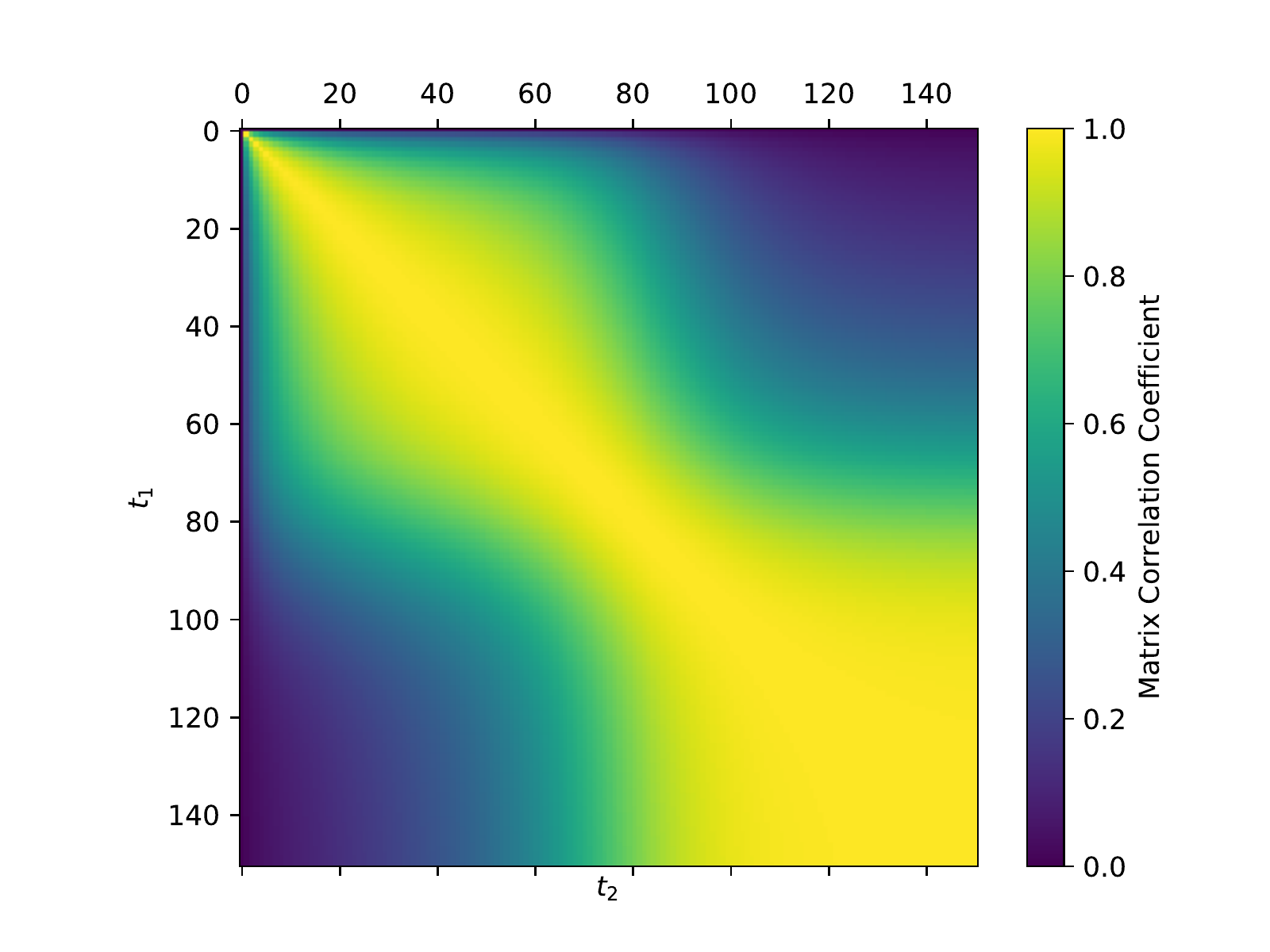}
\caption{Correlation matrix for the number $S$ of susceptible people, i.e.\ $\langle \langle S(t_1) S(t_2) \rangle \rangle/\sqrt(\,\langle \langle S(t_1) S(t_1) \rangle \rangle \langle \langle S(t_2) S(t_2) \rangle \rangle \,)$.}
\labfig{pathcorr}
\end{figure}

 \begin{figure}
\includegraphics[width=\linewidth]{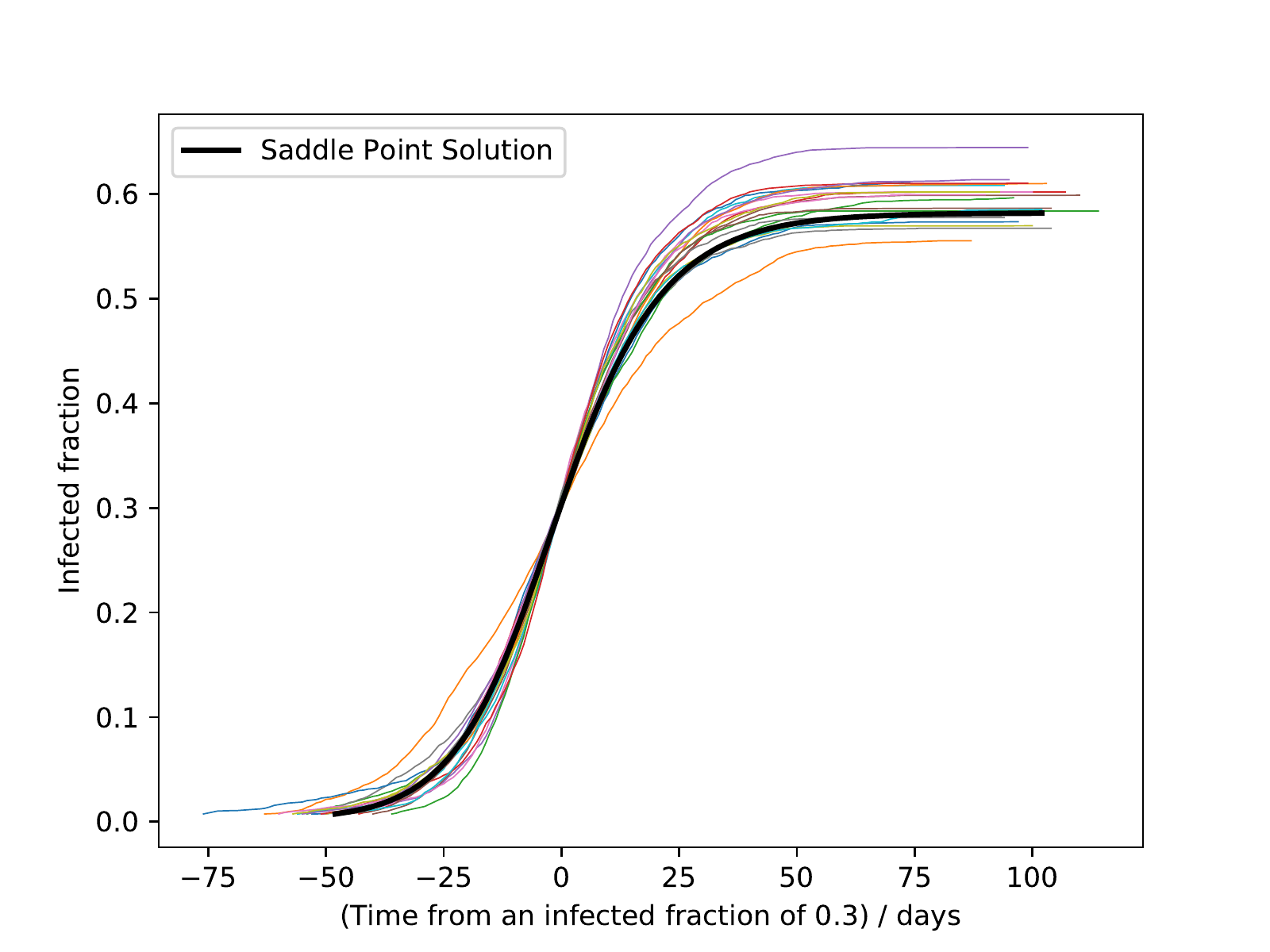}
\includegraphics[width=\linewidth]{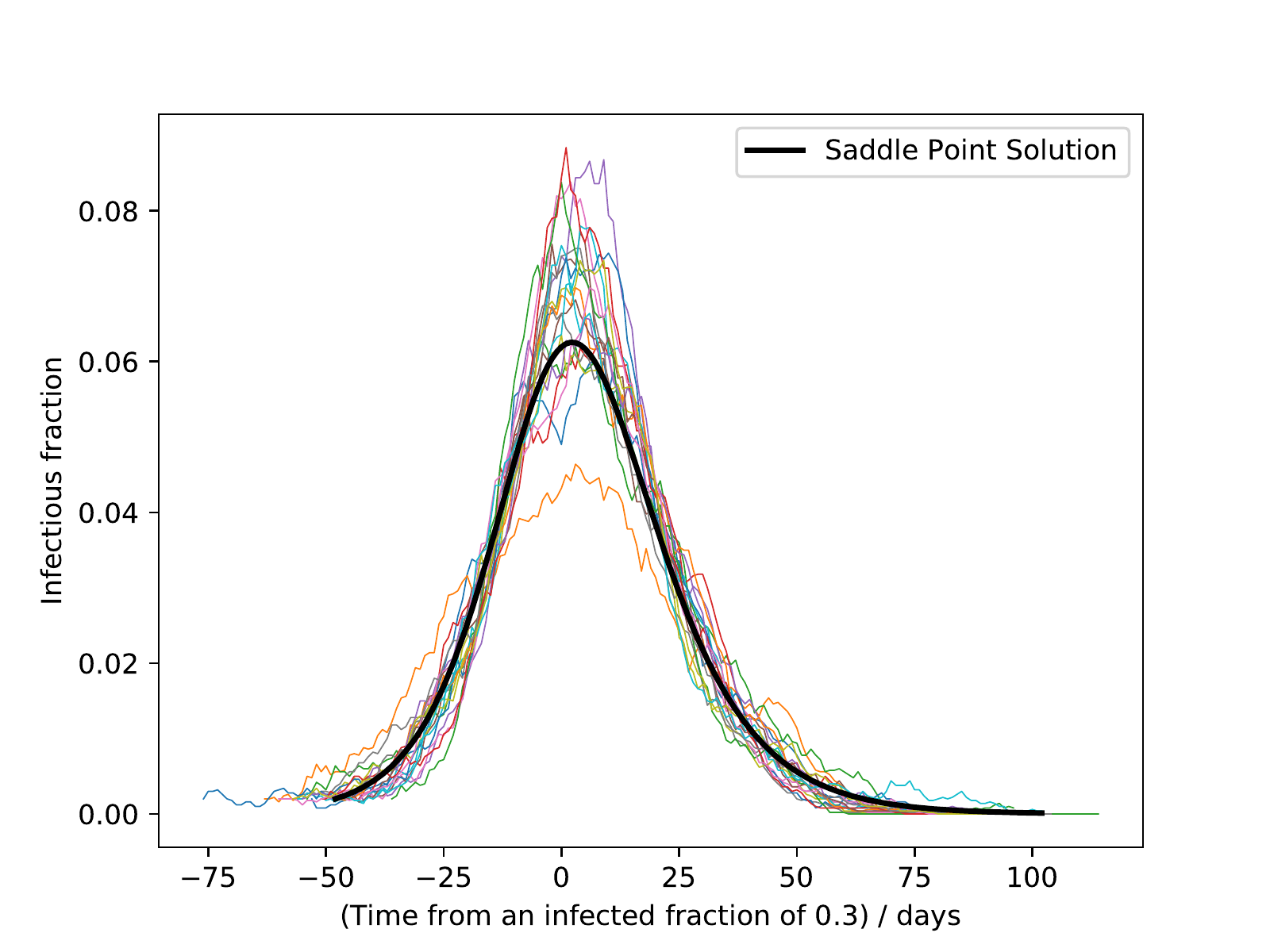}
\caption{Plots illustrating time-displaced infected-fraction-histories (top) and infectious-fraction-histories (bottom) of Fig.\ \reffig{pathsim},  with the displacement chosen in both cases to cause overlap at an infected fraction (nb.\ \emph{not} infectious fraction) of $0.3$.}
\labfig{pathdisp}
\end{figure}

With its ability to calculate unequal-time-correlators, the path integral technique is well-suited to answer questions about relations between quantities at differing times of the epidemic.  For example, in Fig.\ \reffig{pathecl} we compare the number of infectious people on day 30 with the number of susceptible people on day 75.  Remembering $I+T-S$, this requires a number of correlators including $\langle\langle \dt (30) \ds (75) \rangle \rangle$ for example.  The path integral computation captures well the behaviour of the simulated realizations, except for those in which the epidemic dies out early.  As with the discussion around Fig.\ \reffig{pathsim}, it might be possible to use the path integral (looking say in this case at the fraction of the gaussian to the left of $\Delta I(30)=-I(30)$) to estimate how badly it is doing.
\begin{figure}[h!]
\includegraphics[width=\linewidth]{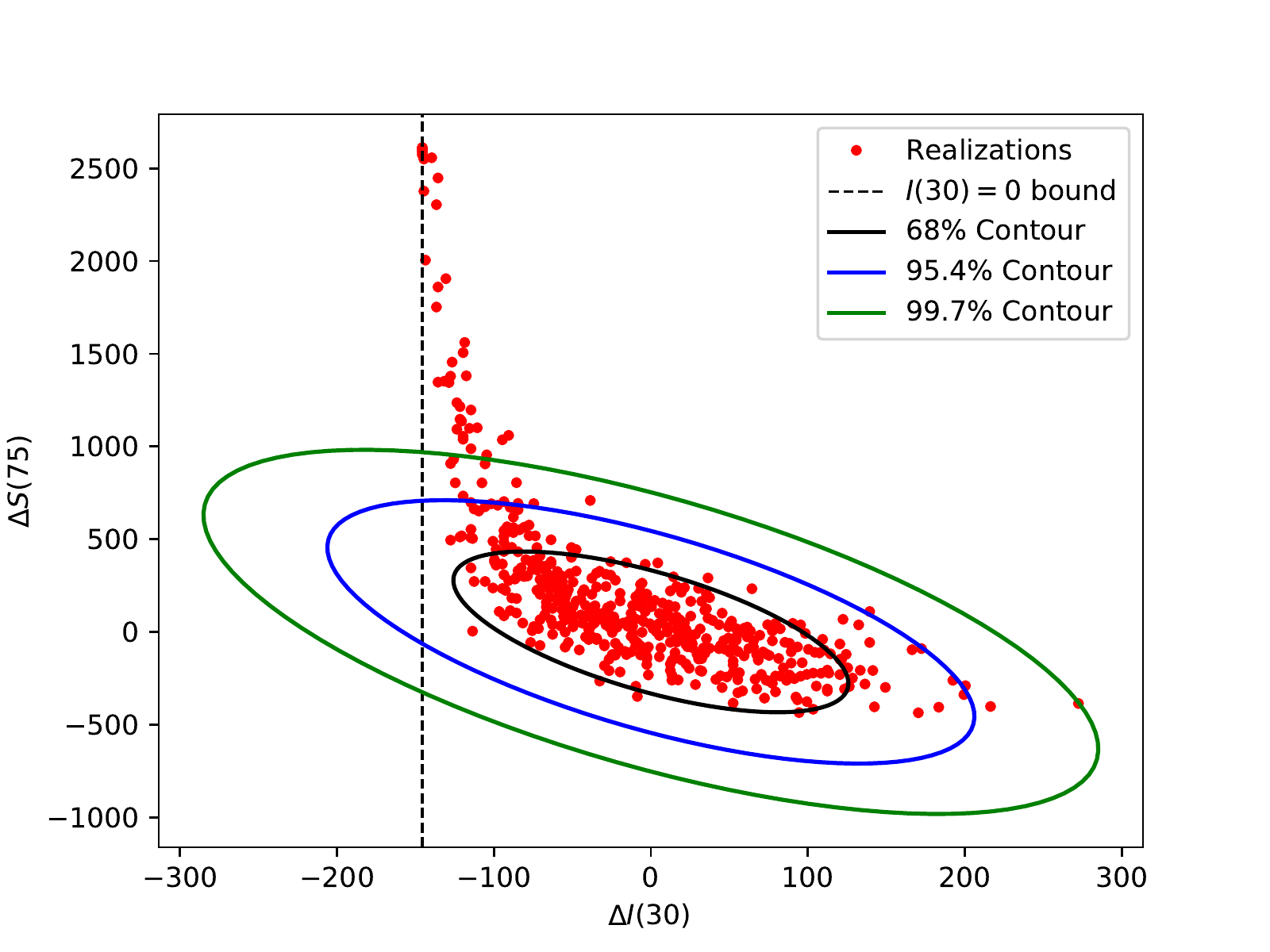}
\caption{Plot showing the scatter around the SIR prediction of the numbers remaining susceptible at $t=75$ days against the numbers of infectious at $t=30$ days from 500 simulations (Nb.\ $I(30)=145.6$ and $S(75)=2349.1$).  The contours are taken from a correlated 2D gaussian distribution with covariances given by the path integral method.  Note the non-gaussian ``tail'' in the simulation results coming from ``failed'' epidemics resulting in a ``pile-up'' of cases near to the limit of zero infectious $\Delta I(30)=-I(30)$, shown by the vertical dashed line.}
\labfig{pathecl}
\end{figure}

\section{Discussion and Conclusions}
\labsec{discon}

In this paper we have seen how it is possible to compute fluctuations around SIR-type models by deterministically solving additional ordinary differential equations instead of by randomly simulating many histories.  While we have only looked at the simplest case here, it would be straightforward to extend this to SEIR models (which add an ``exposed'' stage before infectiousness starts), SEIR models stratified by age or degree of connectivity and so on.  Suitable modifications of the action could handle vaccination schemes and also change the stochastic model of the recovery process if deemed necessary.  Effects of interventions, such as lockdowns, social distancing and isolating, can be incorporated by changing $\beta$ and/or $\gamma$ with time as appropriate.  Some of the equations would then pick up additional time derivative terms\footnote{As happened upon in the construction of simulations to match the base SIR model, it might likely be more realistic to allow $\beta$ and $\gamma$ to change with day and night and presumably by day of the week.}.  With an ``entire history'' view, one can also investigate questions such as how unlikely one would have to be to overwhelm critical care facilities at some period during an epidemic. 
In addition, one could go on to consider if a path integral approach might also be usefully taken to the analysis of network models and to ``real-world'' simulations.

With our example case using only a relatively small population number of 5000, we have seen a number of situations for which a na\"{i}ve ``pure gaussian'' interpretation of the path integral results could lead one into error, typically involving ``early-failing" epidemics.  One can try and creatively reinterpret some of the gaussian outputs as discussed earlier, but it should also be possible to develop other more principled ways of treating this within the path integral framework.  One could add weighting terms to the path integral to implement ``absorbing" boundary conditions at $I=0$ for example, or change the form of the action in certain regimes to better capture the underlying situation\footnote{A perturbative treatment of interaction terms, coming from expanding the action to higher order, as in quantum field theory, would incorporate small amounts of non-gaussianity into the predictions.}.

Although in this work we have focussed on the most likely path going forward from a known initial state and considering the fluctuations around it, there are situations in which one might wish to expand around a path that doesn't necessarily satisfy Eqs.\ \refeq{ds} and \refeq{di} whilst still satisfying Eqs.\ \refeq{els} and \refeq{elt}.  For example, at the end of an epidemic (where $I(\tf)=0$) one might be able to perform accurate serological surveys giving final boundary conditions on $S(\tf)$ also.     Then, if the epidemic was known to start at a given time with a given individual, so $I(0)=1$ and $S(0)=N-1$ there are both initial and final boundary conditions to satisfy.  The extra freedom of Eqs.\ \refeq{els} and \refeq{elt} compared to Eqs.\ \refeq{ds} and \refeq{di} allow a suitable solution to be found for given $\beta$ and $\gamma$.  (By comparing the actions for different $\beta$'s and $\gamma$'s, one then effectively has a likelihood for those parameters solely in terms of the initial and final conditions.) Expanding around such a saddle point, requiring a more complicated version of \refeq{s2} with additional terms now present but with no difference of principle other than a change of boundary conditions for the fluctuations, would then inform one about the spread of variables to be expected in the intermediate stages of that ``constrained'' epidemic.

With a description of fluctuations around a model in hand, one only needs a description of the way chosen data relates to the realization of the model, to be able to go on to construct a likelihood function for the estimation of model parameters and for model comparisons (see also \cite{2020arXiv200509625A}).  Such data might consist of test results, hospitalization and death numbers at various times, with parameters (that could be sampled over and solved for) determining their relation to the underlying epidemic accounting for time delays, incompleteness and so on.  Given the present (as of writing) state of COVID-19 in the UK and elsewhere, to inform lockdown-easing choices it could be very instructive to look at situations in which the effective reproduction number $R_0 \approx 1$ and there is much variability in what might happen next. 

\section*{Acknowledgements}
I thank Christine Gratton for encouragement and for comments on a manuscript of this work. 

\bibliographystyle{unsrt}
\bibliography{bibcovid}

\ifdefined\NOCODE
\else
\appendix
\section{Python Implementation}
\labsec{pyapp}

\lstinputlisting{covid3_stored.py}
\fi
\end{document}